\documentclass[aps,prl,twocolumn,showpacs,superscriptaddress,nofootinbib]{revtex4}  
\usepackage{graphicx}  
\usepackage{dcolumn}   
\usepackage{bm}        
\usepackage{braket}
\usepackage{exscale}                  
\usepackage[intlimits]{amsmath}       
\usepackage{amsfonts}
\usepackage{amssymb,amscd}
\usepackage{color}
\usepackage{hyperref}
\usepackage{subfigure}
\usepackage[normalem]{ulem}

\begin{document}

\title{Universal self-similar scaling of spatial Wilson loops out of equilibrium} 

\author{J.\ Berges}
\email{j.berges@thphys.uni-heidelberg.de}
\affiliation{Institut f\"{u}r Theoretische Physik, Universit\"{a}t Heidelberg, 69120 Heidelberg, Germany}
\author{M.\ Mace}
\email{mark.mace@stonybrook.edu} 
\affiliation{Physics and Astronomy Department, Stony Brook University, Stony Brook, NY 11974, USA}
\affiliation{Physics Department, Brookhaven National Laboratory, Bldg. 510A, Upton, NY 11973, USA}
\author{S.\ Schlichting}
\email{sslng@uw.edu }
\affiliation{Department of Physics, University of Washington, Seattle, WA 98195-1560, USA}

\begin{abstract}	
We investigate strongly correlated non-Abelian plasmas out of equilibrium. Based on numerical simulations, we establish a self-similar scaling property for the time evolution of spatial Wilson loops that characterizes a universal state of matter far from equilibrium. Most remarkably, it exhibits a generalized {\it area law} which holds for sufficiently large {\it ratio} of spatial area and fractional power of time. Performing calculations also for the perturbative regime at higher momenta, we are able to characterize the full nonthermal scaling properties of $SU(2)$ and $SU(3)$ symmetric plasmas from short to large distance scales in terms of two independent universal exponents and associated scaling functions.
\end{abstract}

   
\maketitle

{\it Introduction.} 
Strongly interacting gauge field theories, such as quantum chromodynamics (QCD), have elementary non-perturbative excitations described by Wilson loops~\cite{Wegner,Wilson74}. Such extended objects play an important role also in string theories~\cite{String} or suitable generalizations in formulations of quantum gravity~\cite{LoopQuantumGravity}. In thermal equilibrium the Wilson loop of QCD provides an important means to distinguish the confined ``hadronic'' phase from the deconfined ``quark-gluon plasma'' state~\cite{Aoki:2006we}. Despite the well-established relevance of the Wilson loop for our understanding of fundamental problems in vacuum or thermal equilibrium, much less is known about its significance for dynamical situations out of equilibrium.

In this letter we report on a scaling property of the spatial Wilson loop that characterizes a universal state of QCD matter far from equilibrium. This state provides an important building block for our understanding of the early stages of ultra-relativistic heavy-ion collisions in the limit of sufficiently high energies. In such collisions a nonequilibrium plasma of highly occupied gluons is expected to form~\cite{Lappi,Gelis}, with transient scaling properties~\cite{Berges:2013eia,Berges:2013fga,Kurkela:2015qoa} characterizing the thermalization process of the non-Abelian plasma at weak gauge coupling. However, the notion of occupancies of individual gluons is not gauge invariant and becomes problematic beyond the perturbative high-momentum regime. Since the Wilson loop is gauge invariant, it allows the investigation of non-perturbative ``infrared'' properties of the strongly correlated nonequilibrium system in an unambiguous way~\cite{Berges:2007re,Dumitru:2014,Mace:2016svc}. 

Based on numerical simulations of the plasma's real-time dynamics in the highly excited regime, we establish a self-similar behavior for the time evolution of the spatial Wilson loop. The self-similarity can be fully characterized in terms of a universal scaling exponent and scaling function that are 
time independent. Such universality far from equilibrium is based on the existence of nonthermal fixed points~\cite{Berges:2008wm}, which represent nonequilibrium attractor solutions reached on a time scale much shorter than the asymptotic thermalization time. Since the scaling properties associated with the nonthermal fixed point are insensitive to details of the underlying model and initial conditions, our results provide an important missing piece for the determination of the nonthermal universality classes of non-Abelian gauge theories. We focus here on relativistic non-Abelian plasmas, however, there are important links to similar phenomena in a wide range of applications from cosmology to cold atoms~\cite{Berges:2014bba}. 

While the relevant non-perturbative long-distance or infrared behavior of non-Abelian plasmas can be extracted from Wilson loops, the perturbative scaling properties at higher momenta are well described in terms of quasi-particle excitations and can be inferred from gauge-fixed correlations functions~\cite{Berges:2007re,Berges:2008zt,Schlichting:2012es,Kurkela:2012hp,Mace:2016svc}. Combining both, we establish the full nonthermal scaling properties of (statistically) homogenous and isotropic Yang-Mills plasmas from short to large distance scales. To this end, we also extend previous calculations for the $SU(2)$ gauge group to the case of an $SU(3)$ gauge symmetry underlying QCD. Our results reveal a rather large universality class that is even insensitive to the symmetry group of $SU(2)$ versus $SU(3)$.\\ 

{\it Nonequilibrium Wilson loop.} Wilson loop operators $W$ transport a (color-) electrically charged particle all the way around a closed loop in space-time. Specifically, for a particle charge in the fundamental representation of the non-Abelian $SU(N_c)$ gauge group with $N_c$ colors, the color-averaged transport along a closed line $\mathcal{C}$ is represented by the trace of a path-ordered ($\mathcal{P}$) exponential of the gauge field operator $\mathcal{A}_\mu(x)$~\cite{Wilson74},
\begin{equation}
W  =  \frac{1}{N_c} \mathrm{Tr}\, \mathcal{P} e^{i g \oint_\mathcal{C} d x^\mu \mathcal{A}_\mu(x) } \, ,
\label{eq:WilsonLoop}
\end{equation}  
where $g$ denotes the gauge coupling, and $x^\mu$ are the space-time coordinates with \mbox{$\mu = 0$--$4$}.  

In vacuum or thermal equilibrium the closed curve $C$ is either taken to include the time direction, in which case the time variable is analytically continued to imaginary values, or runs along spatial directions only. The latter is called a spatial Wilson loop. In general, out of equilibrium the time variable may not be continued to imaginary values and we will consider spatial Wilson loops only. The theory is then regularized on a lattice, where the Wilson loop involves products of lattice link variables describing the gauge degrees of freedom~\cite{Wilson74}. 

In equilibrium at zero temperature the expectation value of the Wilson loop $\langle W \rangle_{\mathrm{eq}}$ in the pure gauge theory (without dynamical quarks) decreases exponentially with the area $A$ as 
\begin{equation}
-\log \langle W \rangle_{\mathrm{eq}}  = \sigma_{\mathrm{eq}} \, A
\label{eq:arealaw} 
\end{equation}
for sufficiently large $A$. Specifically, for the  temporal-spatial Wilson loop with imaginary times such an {\it area-law} behavior characterizes confinement, and the associated equilibrium string tension $\sigma_{\mathrm{eq}}$ describes the linear rise of the static quark--anti-quark potential for large spatial separations. At zero temperature, spatial Wilson loops show the same area-law behavior as their temporal-spatial counterparts. However, for spatial Wilson loops this behavior persists even in the deconfined high-temperature phase, where it reflects non-perturbative gauge-field correlations~\cite{Gross:1980br}. 

\begin{figure}[t]
	\centering
	\includegraphics[width=\columnwidth]{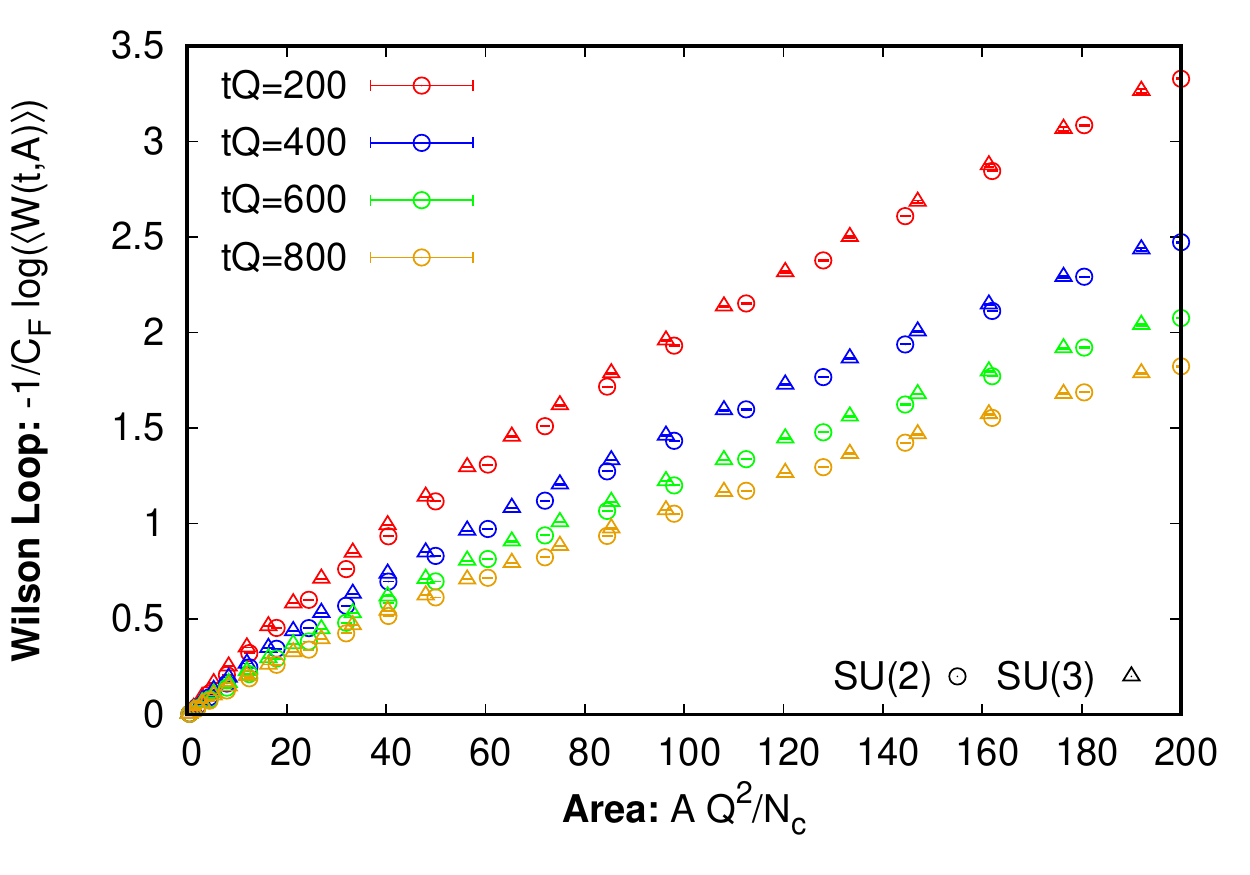}
	\caption{The logarithm of the spatial Wilson loop as a function of the area at different times for gauge groups $SU(N_c)$ with $N_c=2$ (circles) and $N_c=3$ (triangles). Rescaling with the Casimir color factor $-1/C_F(N_c)$ (see main text) leads to very similar results both for two and three colors.
	}
	\label{fig:FIG0}
\end{figure}

Since thermal equilibrium is time-translation invariant, the expectation value $\langle W \rangle_{\mathrm{eq}}$ is a constant in time.  In contrast, for the highly excited nonequilibrium plasma the expectation value $\langle W \rangle(t,A)$ explicitly depends on time $t \equiv x^0$~\cite{Berges:2007re,Dumitru:2014,Mace:2016svc}. A self-similar behavior of the nonequilibrium spatial Wilson loop is described in terms of a universal scaling exponent $\zeta$ and scaling function $w$ as
\begin{equation}
-\log \langle W \rangle (t,A) = w\left((t/t_0)^{-\zeta} A Q^2\right) \, ,
\label{eq:scalingW}
\end{equation}
where $t$ and $A$ are measured in units of a suitable reference time scale $t_0$ and momentum scale $Q$, respectively. Instead of separately depending on time and spatial area of the loop, in a self-similar regime the dynamics only depends on the product of the area and some (fractional) power of time. Such a non-trivial behavior requires a significant loss of information about the microscopic parameters of the underlying system, from which universality originates.\\

{\it Self-similarity and area law scaling at large distances.}
Motivated by the Color Glass Condensate picture of nucleus-nucleus collisions, we consider as an initial condition a nonequilibrium state with energy density $\epsilon \sim Q^4/g^2$ describing an over-occupied gluonic state with characteristic momentum $Q$~\cite{Lappi,Gelis}. Details of the initial conditions are found to become irrelevant on a short time scale $t Q \sim \mathcal{O}(1)$, as demonstrated previously in $SU(2)$ simulations of perturbative quantitives in Refs.~\cite{Schlichting:2012es,Kurkela:2012hp,Mace:2016svc}. While $Q$ is taken to be sufficiently large such that the ``running'' gauge coupling $g(Q)$ is weak due to the phenomenon of asymptotic freedom, the system is strongly correlated because of the non-perturbatively large energy density. 

In this case the nonequilibrium quantum dynamics can be accurately mapped onto a classical-statistical problem, involving sampled solutions of classical Yang-Mills equations for inhomogeneous gauge fields. The latter can be rigorously solved using real-time lattice simulation techniques following Refs.~\cite{Berges:2007re,Berges:2008zt,Schlichting:2012es,Kurkela:2012hp,Mace:2016svc} for the case of a non-expanding plasma~\cite{lattices}. The description reproduces the underlying quantum dynamics at sufficiently early times and breaks down at $tQ\sim g^{-7/2}$ \cite{Schlichting:2012es,Kurkela:2012hp}, when the occupation numbers of typical perturbative momentum modes become of order one such that genuine quantum corrections start playing an important role. 
Since the value of the gauge coupling drops out of the classical-statistical dynamics, the precise value of $g$ merely sets the time scale for the range of validity of our results~\cite{Berges:2013fga}. 

\begin{figure}[t]
	\centering
	\includegraphics[width=\columnwidth]{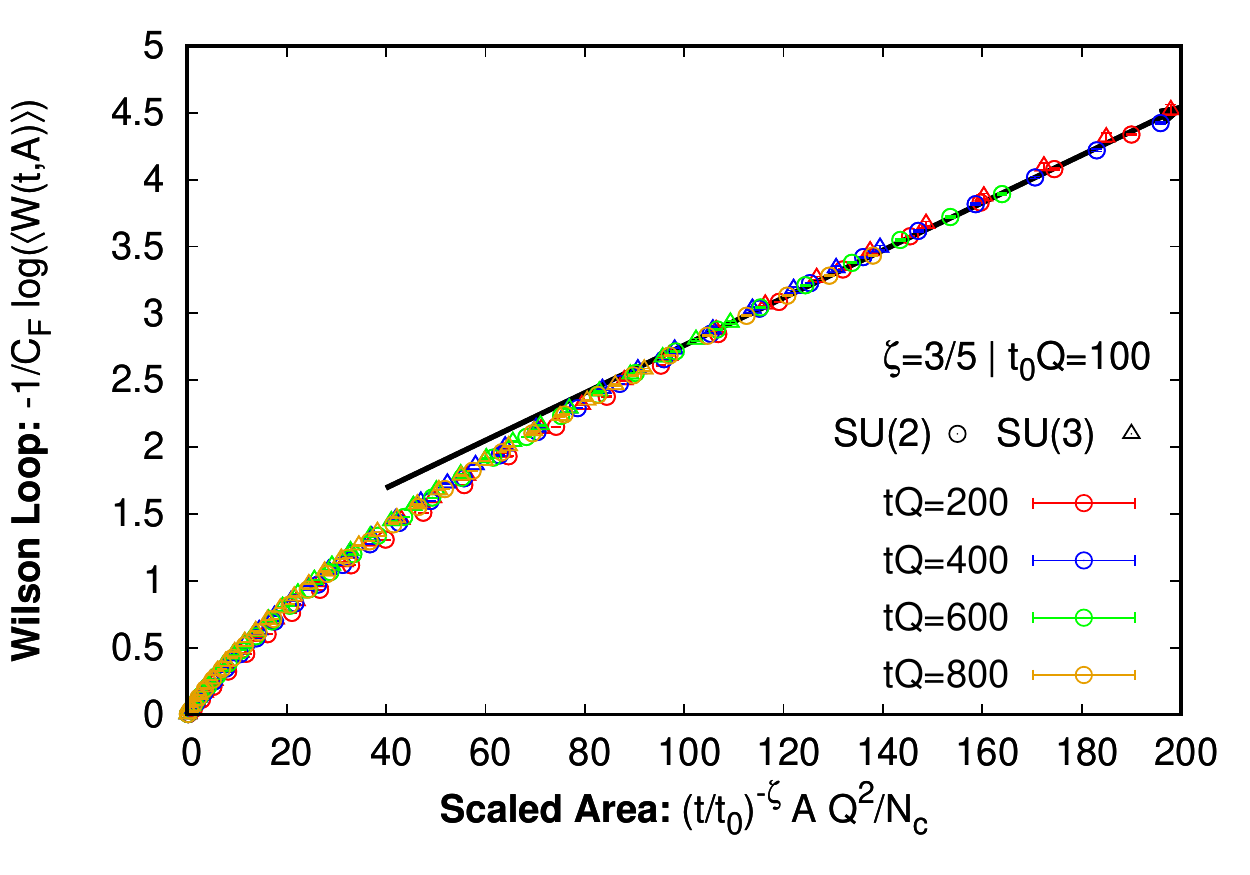}
	\caption{The same as in Fig.~\ref{fig:FIG0}, however, now as a function of the time-rescaled area $\sim t^\zeta A$ with scaling exponent $\zeta = 3/5$. The collapse of the data onto a single curve demonstrates a remarkable level of self-similarity across times, areas, and gauge group. 
	}
	\label{fig:FIG1}
\end{figure}

By virtue of the short-distance expansion for small $A$,
\begin{equation} 
W(t,A)\simeq1-\frac{g^2 }{6 N_c}  A^2 \epsilon_{B}(t) \, ,
\label{eq:WsmallA}
\end{equation} 
where $\epsilon_{B}(t)$ denotes the (time-dependent) color-magnetic energy density~\cite{epsB}, the Wilson loop is expected to approach unity for small enough areas. However, in the case of a possible area-law behavior for large areas $A \gg 1/Q^2$ and sufficiently late times $t \gg 1/Q$ the Wilson loop becomes significantly smaller than unity and decreases monotonically as a function of $A$ such that $-\log \langle W \rangle (t,A) \sim A$.

In Fig.~\ref{fig:FIG0} we present results for the nonequilibrium Wilson loop at different times as a function of the spatial area. Shown is $-\log \langle W \rangle (t,A)$ which approaches zero for $A=0$ and is seen to rise monotonically with the area for $A \gg 1/Q^2$. We display results for both $SU(2)$ and $SU(3)$ gauge groups. After taking into account the Casimir color factors, normalizing the data points with $C_F=(N_c^2-1)/2N_c$ discloses a very similar behavior for $N_c=2$ and $N_c=3$.

Fig.~\ref{fig:FIG1} shows the same data as Fig.~\ref{fig:FIG0}, but now as a function of the {\it rescaled} spatial area $\sim t^{-\zeta} A$ with a time scaling exponent whose numerical fit value suggests \mbox{$\zeta=3/5$}.
Most remarkably, the various sets of data points at different times for $A \gg 1/Q^2$ are found to collapse onto a single curve to very good accuracy. This provides a striking demonstration of the self-similar scaling behavior (\ref{eq:scalingW}). 

Based on the self-similarity observed, we can obtain a precise estimate of the scaling exponent $\zeta$ in (\ref{eq:scalingW}) using a statistical $\chi^2$-analysis as described in Ref.~\cite{Berges:2013fga}. Performing the analysis separately for $N_c=2$ and $N_c=3$, we obtain
\begin{eqnarray}
SU(2):&&\quad \zeta = 0.603 \pm 0.005~(\chi^2) \pm 0.004~(sys.) \, , \nonumber \\
SU(3):&&\quad \zeta = 0.604 \pm 0.004~(\chi^2) \pm 0.005~(sys.) \, 
, \label{eq:zeta}
\end{eqnarray}
where the $\chi^2$-error estimate in the first parentheses is associated to the quality of the scaling for a fixed range of areas and times, and the systematic uncertainties given in the second parentheses are estimated by varying the window in $A$ and $t$ in the analysis. Despite the different structure of the $SU(2)$ and $SU(3)$ gauge group, the respective infrared scaling exponents are found to agree well with each other within errors. 

The behavior of the spatial Wilson loop for large areas reflects the long-distance or ``infrared'' properties of the strongly correlated system. Similar to the large-distance behavior of the spatial Wilson loop in a high-temperature equilibrium plasma~\cite{Gross:1980br}, our data clearly indicates the approach to an area law, which is illustrated by the straight line in Fig.~\ref{fig:FIG1}. However, since the area-law behavior occurs in the self-similar regime of the nonequilibrium evolution, we find that in this case a generalized scaling behavior
\begin{equation}
-\log \langle W \rangle (t,A) \sim t^{-\zeta} A \, 
\label{eq:noneqarealaw}
\end{equation}
holds for a sufficiently large {\it ratio} of spatial area and fractional power of time. Since the scaling exponent $\zeta$ is positive, data points describing larger areas and later times map onto corresponding sets of data points for smaller areas and times. Stated differently, to observe an area law one has to probe larger and larger areas the later the time becomes. In this regime, we may also use (\ref{eq:noneqarealaw}) to define a time-dependent spatial string tension~\cite{Berges:2007re,Dumitru:2014,Mace:2016svc}
\begin{equation}
\sigma(t) = -\frac{\partial \log \langle W \rangle (t,A)}{\partial A} \sim t^{-\zeta} 
\label{eq:spatialstringtension}
\end{equation} 
that is consistent with a previous result~\cite{Mace:2016svc} obtained in the context of sphaleron transitions out of equilibrium.

The area law of the nonequilibrium Wilson loop is only observed at sufficiently large ratio $\sim t^{-\zeta} A$, and Fig.~\ref{fig:FIG1} shows significant deviations to the corresponding straight line for smaller ratios. Since the area law is related to a non-perturbative infrared scale given by the spatial string tension $\sigma$, one may expect a different scaling behavior at shorter length scales where no string tension can be infered. However, Fig.~\ref{fig:FIG1} indicates that the self-similar scaling even holds somewhat beyond the area-law regime. We find that the same exponent $\zeta$ that characterizes the asymptotic scaling of the string tension describes the data well down to $(t/t_0)^{-\zeta} A Q^2/N_c \gtrsim 10$.\\ 

{\it Self-similar scaling in the perturbative regime.} We emphasize that for $A Q^2 \lesssim {\mathcal O}(1)$ there are clear deviations from the self-similar infrared scaling (\ref{eq:scalingW}) observed, which is also expected from the expansion (\ref{eq:WsmallA}) of the Wilson loop for small areas. While the Wilson loop allows one to extract the relevant long-distance properties in a gauge-invariant way, it is less suitable to visualize the detailed short-distance or ultraviolet properties. Besides gauge invariant observables, based e.g. on the energy-momentum tensor, also gauge-fixed quantities provide a valid description for the perturbative higher momenta at weak gauge coupling $g$. Since the gluon distribution function $f(t,p)$ as a function of spatial momentum $p$ and time $t$ has a direct correspondence in kinetic theory, it is typically employed to characterize perturbative scaling properties. The distribution can be extracted from equal-time correlation functions of the gauge fields $\langle \mathcal{A}^{*}_\nu(t,p) \mathcal{A}_\mu(t,p) \rangle$ projected on the transverse polarizations in Coulomb gauge~\cite{Berges:2013fga}. 

\begin{figure}[t]
	\centering
	\includegraphics[width=\columnwidth]{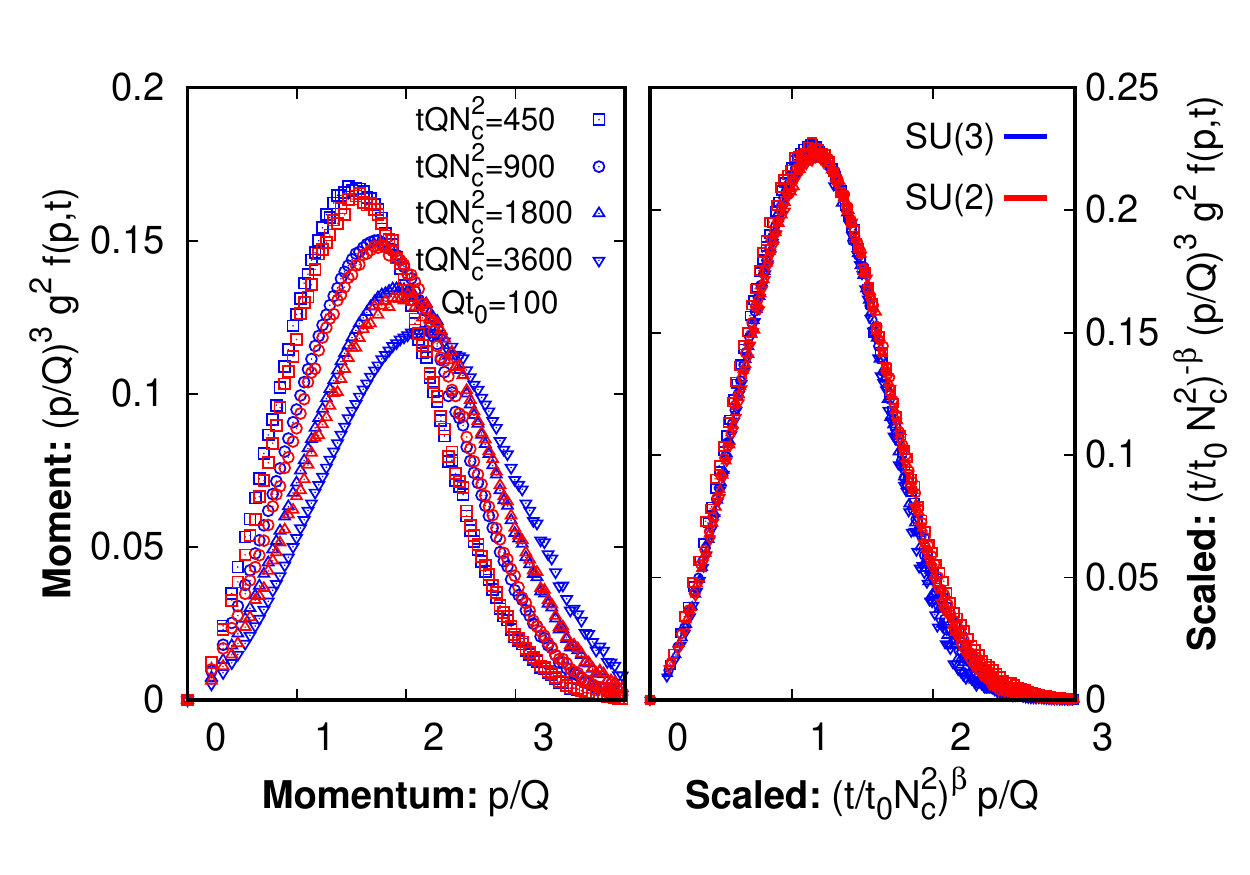}
	\caption{Left: The third moment of the single particle distribution $\sim p^3 f(p,t)$, which is sensitive to perturbative momenta, at different times $ (t/t_0) N_c^2$ for $N_c=2$ and $N_c=3$. Right: The rescaled data for $\beta=-1/7$ collapses onto a single curve demonstrating self-similarity in the perturbative regime.}
	\label{fig:FIG2}
\end{figure}

Self-similarity of the distribution function at higher momenta amounts to 
\begin{eqnarray}
f(t,p) = (t/t_0)^{4\beta}\, f_S\left((t/t_0)^\beta p/Q \right)\, ,
\label{eq:selfsimilarf}
\end{eqnarray}
where the time-dependent normalization $\sim t^{4\beta}$ multiplying the time-independent scaling function $f_S$ arises because of energy conservation, implying that the energy density \mbox{$\epsilon\sim\int d^3p\, p f(t,p) = \mathrm{const}$}. The universal scaling exponent $\beta$ together with the fixed point distribution function $f_S$ characterize the perturbative scaling regime. 

The left panel of Fig.~\ref{fig:FIG2} shows data for the third moment $\sim p^3 f(t,p)$ of the distribution as a function of momentum $p$ at different times $t/t_0 \gg 1$. The figure presents results for $N_c=2$ and $N_c=3$. One observes that the different data sets lie remarkably well on top of each other for given normalized times $N_c^2 t/t_0$. Moreover, if we rescale momenta according to (\ref{eq:selfsimilarf}) with
$\beta = -1/7$
all data sets at different times collapse onto a single time-independent curve as demonstrated in the right panel of Fig.~\ref{fig:FIG2}. For $N_c= 2$ the perturbative scaling behavior has been observed   previously~\cite{Schlichting:2012es,Kurkela:2012hp}
and the value $\beta=-1/7$ is taken from an effective kinetic theory analysis in the perturbative regime~\cite{York:2014wja}. 
Performing the numerical analysis separately for $N_c=2$ and $N_c=3$, we obtain
\begin{eqnarray}
SU(2):&&\quad \beta = -0.145 \pm 0.017~(\chi^2) \pm 0.002~(sys.) \, , \nonumber \\
SU(3):&&\quad \beta = -0.141 \pm 0.020~(\chi^2) \pm 0.002~(sys.) \, 
. \quad\label{eq:beta}
\end{eqnarray}
The fixed point distribution $f_S$ has a universal shape, and the results indicate a rather large universality class that is also insensitive to the symmetry group of $SU(2)$ versus $SU(3)$.  

\begin{figure}[t]
	\centering
	\includegraphics[width=\columnwidth]{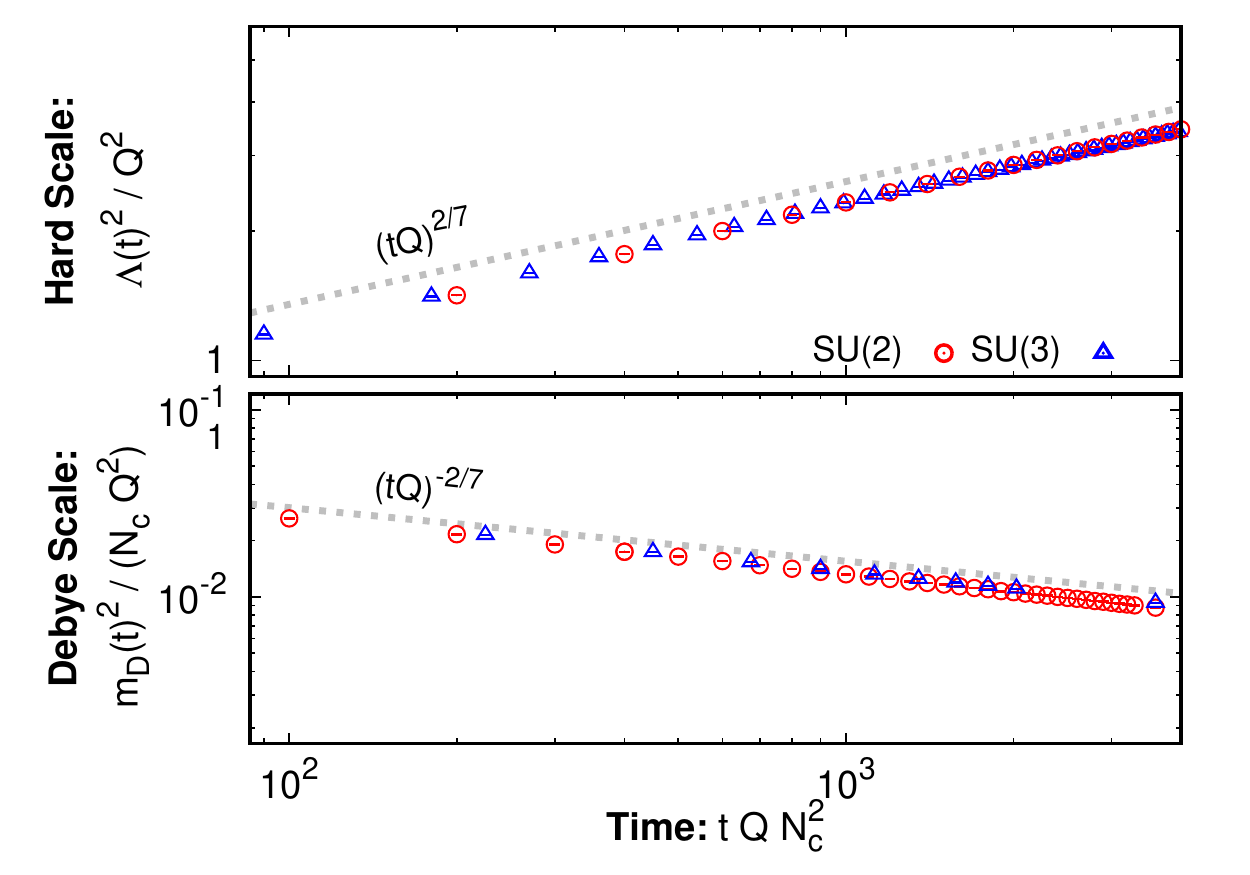}
	\caption{The perturbative hard scale (top) and Debye scale (bottom) as a function of time $t Q N_c^2$ for $N_c = 2$ and $N_c = 3$. Both the two- and three-color cases clearly exhibit the same scaling $\Lambda^2(t) \sim t^{2\beta}$ and $m_D^2(t) \sim t^{-2\beta}$ for $\beta=-1/7$.
	}
	\label{fig:FIG3}
\end{figure}

In the perturbative regime one expects a hierarchy of scales, which appear at different orders of the weak coupling $g$. In an equilibrium plasma at temperature $T$, the ``hard'' momenta are of order $T$, while the color electric and magnetic screening scales are of order $gT$ and $g^2 T$, respectively.
In the nonequilibrium plasma, we characterize the typical momenta $\Lambda(t)$ of hard excitations in terms of a local gauge invariant operator definition constructed from a ratio of covariant derivatives of the field strength tensor and the field strength itself~\cite{Berges:2013fga,Kurkela:2012hp}. Expressed perturbatively, the hard scale is defined as the ratio of moments of the single particle distribution~\cite{Schlichting:2012es,Kurkela:2012hp}
\begin{equation}
\Lambda^2(t) \simeq \frac{2}{3}\frac{\int d^3p\, p^3 f(t,p)}{\int d^3p\, p\, f(t,p)}
\, \sim \, t^{-2 \beta} \, ,
\label{eq:hardscale}
\end{equation}
which explicitly shows the scaling of this quantity with the exponent $\beta$. Here the last equality is obtained from inserting (\ref{eq:selfsimilarf}) into the momentum integral of (\ref{eq:hardscale}). Numerical results are presented in the upper panel of Fig.~\ref{fig:FIG3}, which clearly exhibits the same value of the scaling exponent $\beta$ both for two and three colors. 

At softer momenta, the Debye scale $m_D$ is related to the electric screening scale in plasma. The leading perturbative contribution may be expressed in terms of the distribution function as~\cite{Schlichting:2012es,Kurkela:2012hp}  
\begin{equation}
m_D^2(t) = 4 g^2 N_c \int\! \frac{d^3p}{(2\pi)^3} \frac{f(t,p)}{p}
\, \sim \, t^{2 \beta} \, .
\label{eq:Debye}
\end{equation}
Its scaling behavior is demonstrated in the lower panel of Fig.~\ref{fig:FIG3}. 

Finally, the soft magnetic screening scale in the plasma may be related to the spatial string tension or $\sim \sqrt{\sigma}$. Following a naive power counting, one may be tempted to argue from (\ref{eq:hardscale}) and (\ref{eq:Debye}) that $\sqrt{\sigma}$ should behave \mbox{$\sim t^{3\beta}$}. However, this scale concerns the deep infrared where typical occupancies are of order $f \sim 1/g^2$ in our case, such that a perturbative description is not valid in this regime. Instead, we find according to (\ref{eq:spatialstringtension}) that $\sqrt{\sigma} \sim t^{-\zeta/2}$. With the values (\ref{eq:zeta}) and (\ref{eq:beta}) we note that $-3\beta$ is more than $40\%$ larger than $\zeta/2$, which is well beyond our statistical uncertainties.\\

{\it Conclusions.} Nonthermal fixed points provide an important means to classify and describe the dynamical evolution of strongly correlated systems out of equilibrium. While perturbative scaling properties at high momenta can be understood in terms of self-similar scaling of the single-particle distribution, the notion of quasi-particles with a well-defined momentum becomes inappropriate at soft momenta. Instead, the non-perturbative long-distance behavior is well captured by the elementary excitations of extended objects described by gauge-invariant Wilson loops. Combining both descriptions allows us to establish the full nonthermal scaling properties of the plasma from short to large distance scales. Performing an unprecedented numerical effort in this respect, we are able to characterize the self-similar scaling properties by two independent universal exponents, $\zeta$ in the non-perturbative infrared and $\beta$ in the perturbative ultraviolet regime, and associated scaling functions. We find a remarkable universality between $SU(2)$ and $SU(3)$ Yang-Mills plasmas, which exhibit the same characteristic scaling behavior far from equilibrium even in the deep infrared. 
In view of the significant differences in their thermal equilibrium critical properties, where the $SU(2)$ symmetric theory exhibits long-distance scaling in the Ising universality class and the $SU(3)$ theory is discontinuous at the thermal phase transition, our results point to a rather large universality class for the nonequilibrium scaling phenomenon. Since universal properties are independent of the details of the underlying microscopic system, and the nonthermal behavior is to some extent even insensitive to the symmetry group, this opens the possibility of unexpected links between seemingly disparate physical systems far from equilibrium.\\

{\it Acknowledgments.} We thank N.~Mueller, L.~D.~McLerran, J.~Pawlowski, R.~Pisarski, S.~Sharma, N.~Tanji and R.~Venugopalan for discussions. This work started at the CERN workshop ``The Big Bang and the little bangs - Non-equilibrium phenomena in cosmology and in heavy-ion collisions'', https://indico.cern.ch/event/472353/. This work is part of and supported by the DFG Collaborative Research Centre ``SFB 1225 (ISOQUANT)'', and is supported in part by the U.S.\ Department of Energy under Grants No.\ DE-SC0012704 and DE-FG88-ER40388 (M.M.), DE-FG02-97ER41014 (S.S.) and through the BEST collaboration (M.M.). Numerical calculations used resources of the National Energy Research Scientific Computing Center, a DOE Office of Science User Facility supported by the Office of Science of the U.S.\ Department of Energy under Contract No.\ DE-AC02-05CH11231.


\begin{thebibliography}{99}
\bibitem{lattices}
All data shown are obtained on $128^3$ lattices with spatial
lattice spacing $Qa = 0.5$ and time step $δtQ = 0.01$.
\bibitem{epsB}
In terms of the spatial components $i,j=1,2,3$ of the field-strength tensor $F^a_{ij}[{\mathcal A}]$ with color index $a=1,.., N_c^2-1$ the magnetic energy density is $\epsilon_{B}=\frac{1}{4} F^{a}_{ij}F^{ij}_{a}$ \cite{Wilson74}, where summation over repeated indices is implied.
\end{thebibliography}
\end{document}